\documentclass[12pt,a4paper,aps,preprint,superscriptaddress,nofootinbib]{revtex4-1}

\usepackage{graphicx}
\usepackage{cancel}
\usepackage{amssymb}
\usepackage{textcomp}
\usepackage{amsmath}
\usepackage{bm}
\usepackage{times}
\usepackage{epsfig}
\usepackage{epstopdf}
\usepackage{color}
\usepackage{multirow}
\usepackage[colorlinks=true, pdfstartview=FitV, linkcolor=blue, citecolor=blue, urlcolor=blue]{hyperref}

\def\lsim{\mathrel{\raise.3ex\hbox{$<$\kern-.75em\lower1ex\hbox{$\sim$}}}}
\def\gsim{\mathrel{\raise.3ex\hbox{$>$\kern-.75em\lower1ex\hbox{$\sim$}}}}

\def\ba{\begin{array}}
\def\ea{\end{array}}

\definecolor{orange}{rgb}{1,0.5,0}

\newcommand{\be}{\begin{equation}}
\newcommand{\ee}{\end{equation}}
\newcommand{\bea}{\begin{eqnarray}}
\newcommand{\eea}{\end{eqnarray}}

\begin{document}

\title{Interpretation of the Galactic gamma-ray excess with the dark matter indicated by $^8$Be and $^4$He anomalous transitions}

\author{Lian-Bao Jia}
\email{jialb@mail.nankai.edu.cn}
\affiliation{
School of Science, Southwest University of Science and Technology, Mianyang 621010, China
}
\author{Tong Li}
\email{litong@nankai.edu.cn}
\affiliation{
School of Physics, Nankai University, Tianjin 300071, China
}

\begin{abstract}

The long-standing Galactic center gamma-ray excess could be explained by GeV dark matter (DM) annihilation, while the DM interpretation seems in tension with recent joint limits from different astronomical scale observations, such as dwarf spheroidal galaxies, the Milky Way halo and galaxy groups/clusters. Motivated by $^8$Be and $^4$He anomalous transitions with possible new interactions mediated by a vector boson $X$, we consider a small fraction of DM mainly annihilating into a pair of on-shell vector boson $X X$ followed by $X \to e^+ e^-$ in this paper. The Galactic center gamma-ray excess is explained by this DM cascade annihilation. The gamma rays are mainly from Inverse Compton Scattering emission, and the DM cascade annihilation could be compatible with joint astrophysical limits and meanwhile be allowed by the AMS-02 positron observation. The direct detection of this model is also discussed.

\end{abstract}


\maketitle

\section{Introduction}
\label{sec:Intro}

The existence of Dark Matter (DM) has been established by substantial cosmological and astronomical observations, but the particle nature of DM is still unknown. An appealing candidate of DM is the Weakly Interacting Massive Particles (WIMPs) arising from various extensions of the Standard Model (SM).
The experimental searches for DM consist of categories such as the direct detection (DD) of possible scattering between DM and SM target materials, the indirect detection (ID) looking for signals of DM annihilation/decay products, and the collider search hunting for signals from DM productions at high energy accelerators.

Although confident WIMP signals are still absent from today's DDs of DM and the present DD experiments set strong constraints on WIMPs with masses $\gtrsim$ 10 GeV~\cite{Cui:2017nnn,Akerib:2016vxi,Aprile:2018dbl,Akerib:2017kat,Xia:2018qgs,Aprile:2019dbj,Amole:2019fdf}, anomalous signals from IDs may shed light on some characters of DM. The long-standing Galactic center GeV gamma-ray excess may be signatures of WIMP annihilations \cite{Goodenough:2009gk,Hooper:2010mq,Hooper:2011ti,Abazajian:2012pn,Gordon:2013vta,Daylan:2014rsa,Calore:2014nla}, and millisecond pulsar is another alternative candidate suggested by point sources \cite{Abazajian:2012pn,Gordon:2013vta,Hooper:2013nhl,Yuan:2014rca,Cholis:2014lta,Petrovic:2014xra,Bartels:2015aea,Lee:2015fea,Buschmann:2020adf}. Recent analysis indicate that WIMP annihilations could provide a dominant contribution to the gamma-ray excess after considering possible point sources~\cite{Leane:2019xiy,Zhong:2019ycb,Leane:2020nmi,Leane:2020pfc}. However, this excess regarded as WIMP annihilations seems in tension with recent joint constraints from the observations of dwarf spheroidal galaxies~\cite{Hoof:2018hyn}, the Milky Way halo~\cite{Chang:2018bpt}, the Galactic center~\cite{Abazajian:2020tww} and galaxy groups/clusters~\cite{Lisanti:2017qlb,Lisanti:2017qoz,Tan:2019gmb}. Can WIMP annihilations still be available to explain the Galactic center gamma-ray excess after considering these joint limits? To relax these tensions, some exotic mechanism may be presumable, e.g. a pair of DM particles annihilation into a pair of on-shell light mediators followed by light mediators decay into $e^+ e^-$ or $\mu^+ \mu^-$~\cite{Calore:2014nla,Liu:2014cma,Kaplinghat:2015gha}, with the spectrum of photon arising from radiative processes. For the slightly better fit provided by the scenario of light mediators decay into $e^+ e^-$, the required effective annihilation cross section is $\sim 0.4 \times 10^{-26}$ cm$^3$/s~\cite{Calore:2014nla} which is smaller than the canonical value ($3 \times 10^{-26}$ cm$^3$/s) of DM annihilation cross section.

To interpret the Galactic gamma-ray excess with DM cascade annihilations, what is the light mediator and which new interaction it carries become crucial questions. The new particle and new interaction may leave some traces in anomalous processes. A new mediator $X$ with the mass $m_X \approx$ 16.8 MeV is indicated by the recent $^8$Be \cite{Krasznahorkay:2015iga} and $^4$He \cite{Krasznahorkay:2019lyl} anomalous transitions ($X$ predominantly decays into $e^+ e^-$). We consider the case in which $X$ is a vector boson with primary axial couplings to light quarks as discussed in Ref.~\cite{Kozaczuk:2016nma}, and interpret the Galactic center gamma-ray excess with $X$ boson being the light mediator. The light $X$ boson could induce the Sommerfeld effect~\cite{Sommerfeld:1931} in DM annihilation at low velocities and suffers from the constraints from dwarf galaxies and the Cosmic Microwave Background. In the following discussion, we thus require a vanishing coupling between $X$ and the GeV scale DM indicated by the Galactic gamma-ray excess. To account for the Galactic gamma-ray excess via the DM cascade annihilation, we consider another new scalar $\phi$ which connects DM and $X$ boson. In this paper, the s-wave process of scalar DM $S$ cascade annihilation is $S S^\ast$ $\to$ $\phi$ $\to$ $X X$ followed by $X \rightarrow e^+ e^-$. The $\phi$'s couplings to SM particles could be very small, and thus the DM of concern is able to evade present stringent DD bounds. Note that the DM candidate here may account only a fraction ($f_\mathrm{DM}$) of the total DM of the Universe, referred as multi-component DM scenario~\cite{Zurek:2008qg,Profumo:2009tb}. In this case, the small effective annihilation cross section required by the Galactic gamma-ray excess can be obtained if $S S^\ast$ contributes a small fraction of the total DM.

The paper is outlined as follows. In Sec.~\ref{sec:Model}, we give a brief description of the $X$ boson with Be and He anomalies. We also describe the DM model in terms of mediators $X$ and $\phi$, and derive the DM annihilation cross section. In Sec.\ref{sec:GCE}, we explain the Galactic center gamma-ray excess and discuss the astrophysical constraints on this model. Sec.~\ref{sec:DD} shows our numerical results for the DD of the DM in this model. Our conclusions are drawn in Sec.~\ref{sec:Con}.

\section{New $X$ boson with anomalies and DM hypothesis framework}
\label{sec:Model}

Recently, anomalies were observed in the electromagnetic transitions of both $^8$Be (the 18.15 MeV $1^+$ state)~\cite{Krasznahorkay:2015iga} and $^4$He (the 21.01 MeV $0^-$ state)~\cite{Krasznahorkay:2019lyl}. These two anomalies may be caused by the same origin, i.e. a new mediator $X$ predominantly decaying into $e^+ e^-$ with the mass $m_X \approx$ 16.8 MeV. The nuclear transitions of $^8$Be were analyzed in Ref.~\cite{Zhang:2017zap} with improved nuclear physics model, although this way cannot explain the anomaly. To account for the $^8$Be anomaly, the spin-parity of the $X$ boson could be $J^P = 1^-$, $1^+$, or $0^-$. The corresponding $X$'s couplings to SM particles of the vector/axial-vector~\cite{Feng:2016jff,Feng:2016ysn,Gu:2016ege,Kozaczuk:2016nma,Feng:2020mbt,Seto:2020jal} and pseudoscalar~\cite{Ellwanger:2016wfe} forms were analyzed in the literatures (see Refs.~\cite{Liang:2016ffe,Fayet:2016nyc,DelleRose:2017xil,Fornal:2017msy,Jiang:2018uhs,Nam:2019osu} for more studies). In addition, the spin-parity of the $X$ boson required by the $^4$He anomaly is $J^P = 1^+$, $0^-$. Thus, if parity is conserved in the two anomalous nuclear transitions of $^8$Be and $^4$He, $X$'s couplings to light quarks could be of axial-vector ($1^+$) or pseudoscalar ($0^-$) forms. Further estimates of the decay widths indicate that a pseudoscalar $X$ for the two anomalies is disfavored~\cite{Feng:2020mbt}.

The $^4$He anomaly was analyzed by the experimental group~\cite{Krasznahorkay:2019lyl} with the $0^-$ state in the electro-magnetically forbidden transition $0^- \to 0^+ + X$. Note that, if the He anomaly is produced by the excited state 0$^+$ in $0^+ \to 0^+ + X$ transition, the spin-parity of the $X$ boson could be $J^P$ = $1^-$ (a vector $X$) for both Be and He anomalies~\cite{Feng:2020mbt}. Here we follow the experimental group and focus on the excited state $0^-$, and thus the spin-parity of the $X$ boson is taken as $J^P$ = $1^+$ for both Be and He anomalies. To account for anomalous $^8$Be and $^4$He transitions, the forms of $X$'s couplings to quarks and charged leptons are taken as that in Ref.~\cite{Kozaczuk:2016nma}
\begin{eqnarray}
\mathcal{L}_{X} \supset -  X_\mu \sum_q  g_q  \bar{q} \gamma^\mu \gamma^5 q  + X_\mu \sum_i \bar{l}_i ( g_i^V  \gamma^\mu + g_i^A  \gamma^\mu \gamma^5 ) l_i  \; ,
\end{eqnarray}
where we assume the $X$ particle predominantly decays into $e^+ e^-$. The $X$ couplings to $u$, $d$ quarks are in the range of $10^{-5} \lesssim |g_{u,d}|  \lesssim  10^{-4}$~\cite{Kozaczuk:2016nma} and the typical value of the coupling parameter $\sqrt{(g_e^V)^2 + (g_e^A)^2}/e \sim  10^{-3}$ (or $\sim 5\times10^{-5}$) is allowed by experiments~\cite{Riordan:1987aw,Alexander:2016aln,Banerjee:2019hmi}. The mean decay length $L$ of the $X$ boson should be $L \lesssim$ 1 cm~\cite{Feng:2016jff}. The current and near future experiments can probe the parameter space of interest for the $X$ boson~\cite{Kozaczuk:2016nma}. It can be tested in other large-energy nuclear transitions, such as the transitions of $^{10}B$ and $^{10}Be$ to their ground state~\cite{Arvieux:1967tqs,Subotic:1978mab}. The search for dark photon $A'$ through $D^*(2007)^0 \to D^0 + A'$ with $A'\to e^+ e^-$ at LHCb can probe the region of the $X$ gauge boson explaining the anomalies. The Mu3e experiment will be sensitive to dark photon masses above 10 MeV. The possible dark photon can be produced via electron scattering off a gas hydrogen target in DarkLight experiment~\cite{Balewski:2014pxa}. The Heavy Photon Search experiment searches for $X\to e^+ e^-$ through a high-luminosity electron beam incident on a tungsten target~\cite{Moreno:2013mja}. The $X$ boson can be produced in the bremsstrahlung reaction of $e^- Z \to e^- Z X$ at NA64~\cite{Banerjee:2019hmi}, and the TREK experiment at J-PARC is expected to look for $K\to \mu\nu (X\to e^+ e^-)$. The $e^+e^-$ colliders have the capacity to search for $e^+ e^-\to X\gamma$. The new $X$ boson may couple to the dark sector and bridge the transition between the SM and the dark sector. The possible $X$ portal DM was investigated in Refs.~\cite{Jia:2016uxs,Kitahara:2016zyb,Chen:2016tdz,Seto:2016pks,Jia:2017iyc,Jia:2018mkc}. The masses of the $X$ portal DM in these models are generally in MeV scale~\cite{Jia:2016uxs,Jia:2018mkc}. In this paper, we focus on interpreting the Galactic gamma-ray excess with new interactions mediated by $X$ boson in DM cascade annihilations.

Next, we consider the framework that a fraction ($f_\mathrm{DM}$) of DM particles consist of a gauge singlet complex scalar $S$ and the main transitions between $S$ and the SM particles occur via another scalar field $\phi$ and a spin-1 mediator $X$. The $X$ boson causes the anomalous transitions of $^8$Be and $^4$He, but it does not directly couple to $S$ to avoid the possible Sommerfeld effect. The real dark field $\phi$ couples to both $S$ and $X$, which mediates the transition between $S S^\ast$ and the SM particles. Besides the kinetic energy terms, the general Lagrangian is given by~\footnote{Suppose $X$ is charged under an extra dark Abelian gauge symmetry, one can introduce the covariant derivative operator $(D_\mu \Phi)^\dagger (D^\mu \Phi)$. After the scalar field gets the vacuum expectation value $\langle\Phi\rangle\to v_D$, the Abelian symmetry is broken and one gets the mass of neutral gauge
boson $X$. The real scalar field $\phi$ in $\Phi=\phi+v_D$ and the last two terms of $X$-$\phi$ couplings in the Lagrangian can be also induced.}
\begin{eqnarray}
-\mathcal{L}_{scalar} &\supset& \mu^2 |H|^2 + \lambda |H|^4 + \mu_\phi |H|^2\phi- {1\over 2}\mu_\phi v^2\phi + \lambda_\phi |H|^2\phi^2+m_\phi^2\phi^2+\lambda_4\phi^4\nonumber \\
&+&\mu_S S S^\ast \phi + \lambda_S S S^\ast \phi^2 +\mu_X X^\mu X_\mu \phi + \lambda_X X^\mu X_\mu \phi^2\; ,
\end{eqnarray}
where $H$ is the SU$(2)_L$ Higgs doublet as $H=(0,h/\sqrt{2})^T$. We assume $S$ is invariant under a global U$(1)$ symmetry in dark sector which eliminates all terms with complex coefficients.
We follow Refs.~\cite{Liu:2014cma,Pospelov:2007mp} to neglect the Higgs portal terms $SS^\ast |H|^2$ and $XX|H|^2$ as the new sector could be sequestered.~\footnote{If the bosonic particles are written as superfields in a SUSY theory at a higher scale, the superpotential does not lead to these Higgs portal terms~\cite{Liu:2014cma}. Also, the sectors can be sequestered if the theory arises from an extra dimension~\cite{Liu:2014cma}. As a result, the Higgs field $H$ does not directly couple to $S$ or $X$, with the absent $SS^\ast |H|^2$ and $XX|H|^2$ terms.}
After the SM Higgs doublet develops a vev ($v$), the electroweak symmetry is broken. The trilinear term $\mu_\phi |H|^2\phi$ is introduced to mix $\phi$ and $H$. We also choose the linear term of $\phi$ as $-{1\over 2}\mu_\phi v^2\phi$ to make sure $\phi$ does not obtain a vev \cite{OConnell:2006rsp,Pospelov:2007mp}. The $\mu_\phi$ term is very small in order to introduce a tiny mixing between $H$ and $\phi$.
The introduction of the trilinear term is equivalent to the case in which the singlet develops a vev spontaneously.
The latter case, however, would suffer from a domain wall problem. The phenomenological results of the two cases are the same. By minimizing the above potential, one arrives at the condition as $\mu^2=-\lambda v^2$. The mass spectrum is obtained as follows
\begin{eqnarray}
\mathcal{M}^2&=&  \left(
  \begin{array}{cc}
    2\lambda v^2 & \mu_\phi v \\
    \mu_\phi v & 2m_\phi^2+\lambda_\phi v^2 \\
  \end{array}
\right) \,.
\end{eqnarray}
The mass eigenstates after diagonalizing the CP-even scalars are
\begin{eqnarray}
\left( \ba{c} h_1  \\ h_2  \ea  \right) &=&\left(
\ba{cc}
\cos\theta & \sin\theta   \\
-\sin\theta & \cos\theta   \\
\ea \right) \left( \ba{c} h  \\ \phi \ea  \right)\,,
\end{eqnarray}
with the masses of $h_1$ and $h_2$ being $M_1$ and $M_2$, respectively. The mixing angle $\theta$ is very tiny with $\sin\theta\ll 1$. We should keep in mind that there may be more particles in the new sector, and the particles playing key roles in transitions between the SM and the dark sector are considered here.

In terms of the mass eigenstates, our parameter inputs can be traded into the scalar masses and the mixing angle as
\begin{eqnarray}
\lambda&=&\frac{1}{2 v^2} \Big( \cos^2\theta M_1^2 + \sin^2\theta M_2^2  \Big) \,,\\
\mu_\phi&=&\sin\theta\cos\theta(M_1^2-M_2^2)/v\;,\\
m_\phi^2&=&{1\over 2}\Big(\sin^2\theta M_1^2 + \cos^2\theta M_2^2 -\lambda_\phi v^2\Big)\;.
\end{eqnarray}
Besides $M_1=125$ GeV and $v\approx 246$ GeV, the free parameters in the physical basis are
\begin{eqnarray}
M_2, \theta, \lambda_\phi, \lambda_4, \mu_S, \lambda_S, \mu_X, \lambda_X\;.
\end{eqnarray}
The relevant cubic Higgs and quartic self-couplings in the physical basis are
\begin{eqnarray}
\lambda_{111}&=&3\Big({c^3_\theta M_1^2\over v}+ 2s^2_\theta c_\theta \lambda_\phi v\Big)\;, \\
\lambda_{1111}&=&3\Big({c^6_\theta M_1^2+c^4_\theta s^2_\theta M_2^2\over v^2}+ 4c^2_\theta s^2_\theta \lambda_\phi + 8s^4_\theta \lambda_4\Big)\;,\\
\lambda_{211}&=&-{s_\theta\over v}\Big(2c^2_\theta M_1^2+c^2_\theta M_2^2-4\lambda_\phi c^2_\theta v^2+2\lambda_\phi s^2_\theta v^2\Big)\;,\\
\lambda_{SS1}&=&s_\theta \mu_S\;, \lambda_{SS2}=c_\theta \mu_S\;, \lambda_{SS11}=2s^2_\theta \lambda_S\;,\lambda_{SS22}=2c^2_\theta \lambda_S\;,
\lambda_{SS12}=2c_\theta s_\theta \lambda_S\;,\\
\lambda_{XX1}&=&2s_\theta \mu_X\;,\lambda_{XX2}=2c_\theta \mu_X\;,\lambda_{XX11}=4s^2_\theta \lambda_X\;,\lambda_{XX22}=4c^2_\theta \lambda_X\;,
\lambda_{XX12}=4c_\theta s_\theta \lambda_X\;,
\end{eqnarray}
with $s_\theta=\sin\theta$ and $c_\theta=\cos\theta$.

To give an explanation of the Galactic gamma-ray excess via the scenario of DM cascade annihilations, here we assume $M_2/2 > M_1/4, m_S \gg m_X$ for simplicity. For scalar DM $S$ the annihilation process is $S S^\ast \to h_1, h_2\to X X$ followed by $X$ decaying into $e^+ e^-$. The annihilation cross section is
\begin{eqnarray}
\sigma  &=& \frac{s^4_\theta \mu_S^2 \mu_X^2}{8 \pi s}  \sqrt{\frac{s-4m_X^2}{s-4m_S^2}} \frac{ s^2/(4m_X^4)-s/m_X^2+3}{(s - M_1^2)^2 + M_1^2 \Gamma_{h_1}^2}\nonumber \\
&+& \frac{c^4_\theta \mu_S^2 \mu_X^2}{8 \pi s}  \sqrt{\frac{s-4m_X^2}{s-4m_S^2}} \frac{ s^2/(4m_X^4)-s/m_X^2+3}{(s - M_2^2)^2 + M_2^2 \Gamma_{h_2}^2} \; , \nonumber \\
\sigma v_r &\simeq& {c^4_\theta \mu_S^2 \mu_X^2\over 16\pi m_S^3}\Big(4{m_S^4\over m_X^4}-4{m_S^2\over m_X^2}+3\Big){\sqrt{m_S^2-m_X^2}\over (M_2^2-4m_S^2)^2 }\; ,
\end{eqnarray}
where $v_r$ is the relative velocity, and $s$ is the squared total invariant mass. The case that the annihilation of DM is away from the resonance is of our concern. In addition, the decay of $h_2$ includes $h_2\to SS^\ast, XX$ for $M_1>M_2/2$ or $h_2\to SS^\ast, XX, h_1h_1$ for $M_2/2>M_1$, and their partial decay widths are
\begin{eqnarray}
\Gamma(h_2\to SS^\ast)&=& {c^2_\theta \mu_S^2\over 16\pi M_2}\sqrt{1-{4m_S^2\over M_2^2}} \;,\\
\Gamma(h_2\to XX)&=& {c^2_\theta \mu_X^2\over 8\pi M_2}\sqrt{1-{4m_X^2\over M_2^2}} \Big({M_2^4\over 4m_X^4}-{M_2^2\over m_X^2}+3\Big) \;,\\
\Gamma(h_2\to h_1h_1)&=& {\lambda_{211}^2\over 32\pi M_2}\sqrt{1-{4M_1^2\over M_2^2}}\;.
\end{eqnarray}

Besides the scalar DM $S$ of our concern, a very similar scenario is the case of vector DM $V$. The corresponding Lagrangian is
\begin{eqnarray}
-\mathcal{L}_{V} &\supset& \mu^2 |H|^2 + \lambda |H|^4 + \mu_\phi |H|^2\phi - {1\over 2}\mu_\phi v^2\phi + \lambda_\phi |H|^2\phi^2+m_\phi^2\phi^2+\lambda_4\phi^4\nonumber \\
&+&\mu_V V V^\ast \phi + \lambda_V V V^\ast \phi^2 +\mu_X X^\mu X_\mu \phi + \lambda_X X^\mu X_\mu \phi^2\; .
\end{eqnarray}
The s-wave annihilation process $V V^\ast \to h_1, h_2\to X X$ followed by $X$ decaying into $e^+ e^-$ could also account for the Galactic center gamma-ray excess as that of scalar DM $SS^\ast$. The phenomenology is the same for our purpose.

\section{The Galactic center gamma-ray excess and other constraints}
\label{sec:GCE}

The particles $S S^\ast$ acquire their relic abundance through cascade annihilation processes. For single-component thermally freeze-out DM with DM annihilations in s-wave, the annihilation cross section of DM is about $2.2 \times 10^{-26}$ cm$^3$/s for DM mass $\gtrsim$ 10 GeV~\cite{Steigman:2012nb}. In fact, one possible DM candidate may account only a fraction $f_\mathrm{DM}$ of the total DM in the Universe. For non-self-conjugate DM $S$, the annihilation cross section is required to be $4.4 \times 10^{-26} / f_\mathrm{DM}$ cm$^3$/s. The present cascade annihilation of $S S^\ast$ may account for the long-standing Galactic center gamma-ray excess, and next we will give an analysis about it.

\subsection{The Galactic center gamma-ray excess}

In the cascade annihilation $S S^\ast \to h_1, h_2 \to X X$ followed by $X$ decaying into $e^+ e^-$, the electron and positron are boosted in the final state. In this paper, we have the mass relation $m_S \gg m_X \gg m_e$ and thus the corresponding energy of $e^+$ and $e^-$ can be as high as $m_S$. Possible gamma-ray signal could be produced via Inverse Compton Scattering (ICS) and bremsstrahlung in DM cascade annihilation. To account for the Galactic center gamma-ray excess peaked around 1-3 GeV, the contribution from ICS emission is crucial and the contribution from bremsstrahlung emission is subdominant.

\begin{figure}[htbp!]
\includegraphics[width=0.45\textwidth]{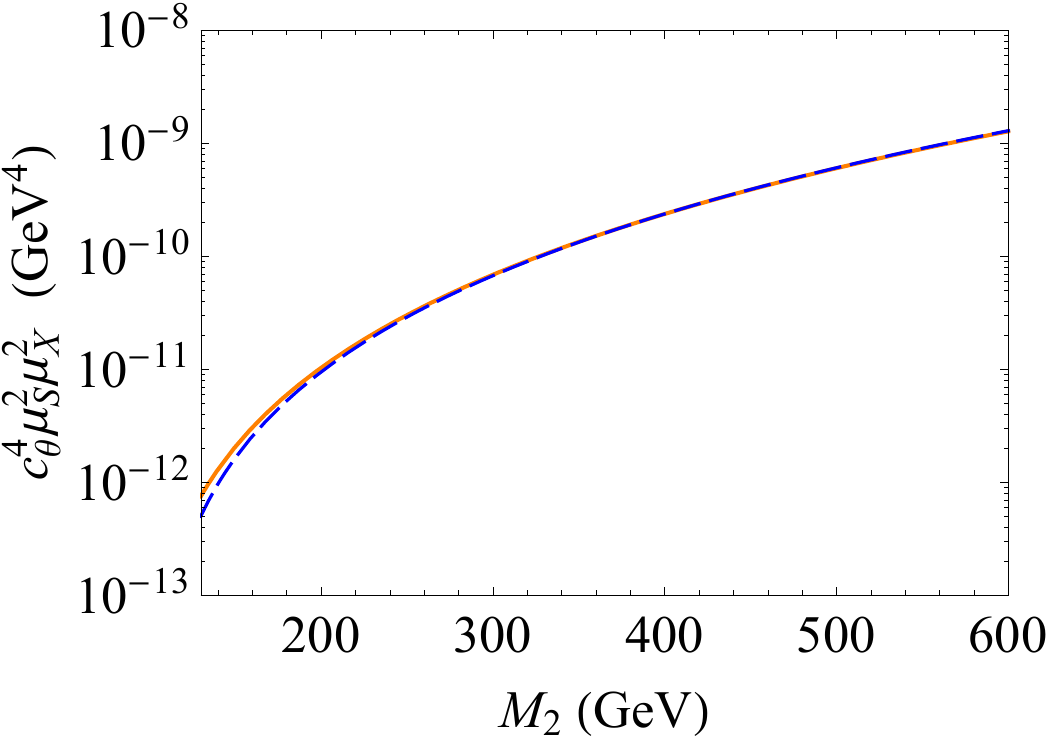} \vspace*{-1ex}
\caption{The coupling parameters as an function of the mediator's mass $M_2$. Here $M_2$ varies in a range of 130$-$600 GeV. The solid and dashed curves are for the WIMP dominant and SIDM dominant cases, respectively.}\label{coupling}
\end{figure}

For the annihilation of $S S^\ast$ with a small fraction $f_\mathrm{DM}$, we assume that the DM spatial distribution follows that of the main DM component. The first case is that the main DM component is WIMP type particles with a generalized Navarro-Frenk-White density profile. In this case the $S$ component with the mass $m_S = 45.7^{+3.4}_{-3.3}$ GeV and the revised annihilation cross section $f_\mathrm{DM}^2 \langle \sigma v_r \rangle /2 = 0.384^{+0.052}_{-0.051} \times 10^{-26}$ cm$^3$/s (the factor 1/2 is for the $S S^\ast$ pair required in annihilation) can fit the GeV excess~\cite{Calore:2014nla}. In the second case the main DM component is self-interacting DM (SIDM), motivated by the small-scale problems. The density of SIDM at the Galactic center could be comparable to or larger than the cold DM predictions when SIDM tracks the baryonic potential \cite{Kaplinghat:2013xca}. In this case the $S$ component with the mass $m_S \sim 50$ GeV and the revised annihilation cross section $f_\mathrm{DM}^2 \langle \sigma v_r \rangle  \sim 6.3 \times 10^{-27}$ cm$^3$/s could fit the GeV excess~\cite{Kaplinghat:2015gha}. Thus, to explain the Galactic gamma-ray excess, we adopt two benchmark points for DM mass and the revised annihilation cross section with [$m_S=45.7$ GeV, $f_\mathrm{DM}^2 \langle \sigma v_r \rangle /2=0.384\times 10^{-26}$ cm$^3$/s] for WIMP dominant case and [$m_S=50$ GeV, $f_\mathrm{DM}^2 \langle \sigma v_r \rangle /2=0.315\times 10^{-26}$ cm$^3$/s] for SIDM dominant case. The corresponding fraction $f_\mathrm{DM}$ can then be derived as
\begin{eqnarray}
 f_\mathrm{DM} \simeq \bigg \{ \begin{array}{cc}
  0.175  \, ,  &  \rm for ~ WIMP ~ dominant \, \\
  0.143  \, , &  \rm for ~ SIDM ~ dominant \,
\end{array} \, .
\end{eqnarray}
We consider the case that the DM annihilation is away from the resonance with $M_2 /2 m_S \gtrsim 1.3$ (see e.g. Ref.~\cite{Jia:2019yhr}). Thus, today's annihilation cross section of $S S^*$ is equal to that at the freeze-out epoch. We can obtain the coupling parameter saturating the annihilation cross section, as an function of the mediator's mass $M_2$ shown in Fig.~\ref{coupling}.

\subsection{Other astrophysical constraints}

As stated in the Introduction, the explanation of the Galactic center gamma-ray excess with DM annihilations needs to be compatible with the constraints from different scale astrophysical observations, such as dwarf spheroidal galaxies, the Milky Way halo and galaxy groups/clusters. Here we will give a brief discussion about whether the cascade annihilation of concern is compatible with joint astrophysical limits.

To account for the 1-3 GeV gamma-ray excess at Galactic center via DM cascade annihilation, the contribution from ICS emission is dominant and that from bremsstrahlung emission is subdominant. The ICS emission is closely related to the distribution of the ambient photon background, and the bremsstrahlung emission is related to the ambient gas densities. Hence, the constraints from dwarf spheroidal galaxies~\cite{Hoof:2018hyn} and the Milky Way halo (regions away from the Galactic plane)~\cite{Chang:2018bpt} are relaxed due to low starlight and gas densities. New likelihood analyses of the Galactic center region~\cite{Abazajian:2020tww} set strong constraints on hadronic components produced by DM annihilations, while the constrains for light lepton components of $\mu^+ \mu^-$, $e^+ e^-$ are relaxed. As the DM of concern here predominantly annihilates in s-wave, the constrains from galaxy groups/clusters~\cite{Lisanti:2017qlb,Lisanti:2017qoz,Tan:2019gmb} with large relative velocity are alleviated.

For DM cascade annihilations accounting for the gamma-ray excess with the contribution mainly from ICS emission, the most severe restriction is from the positron fraction observed by AMS-02. In the case of DM mass in tens of GeV, the effective cross section of DM cascade annihilations today indicated by AMS-02~\cite{Elor:2015bho} should be $\lesssim 1 \times 10^{-26}$ cm$^3/s$, and it can be achieved by a small fraction $f_\mathrm{DM}$ of DM participating in the cascade process (the effective annihilation cross section today is proportional to $f_\mathrm{DM}^2$). For the WIMP dominant case, the upper limit of the revised annihilation cross section set by AMS-02 is about $0.8\times 10^{-26}$ cm$^3$/s with $m_S\simeq 45.7$ GeV~\cite{Elor:2015bho}, and thus the benchmark point [45.7 GeV, $0.384\times 10^{-26}$ cm$^3$/s] is allowed by the AMS-02. For the SIDM dominant case, the fact that the SIDM tracks the baryonic potential can raise the density of SIDM at the Galactic center~\cite{Kaplinghat:2013xca}. In this case, the benchmark point [50 GeV, $0.315\times 10^{-26}$ cm$^3$/s] could fit the GeV excess and is allowed by the AMS-02 constraint~\cite{Kaplinghat:2015gha}. Thus, the DM cascade annihilation of concern can give an interpretation on the Galactic gamma-ray excess and meanwhile is compatible with the joint astrophysical constraints mentioned above.

\section{Direct detection of DM}
\label{sec:DD}

\begin{figure}[htbp!]
\begin{center}
\includegraphics[scale=1,width=7cm]{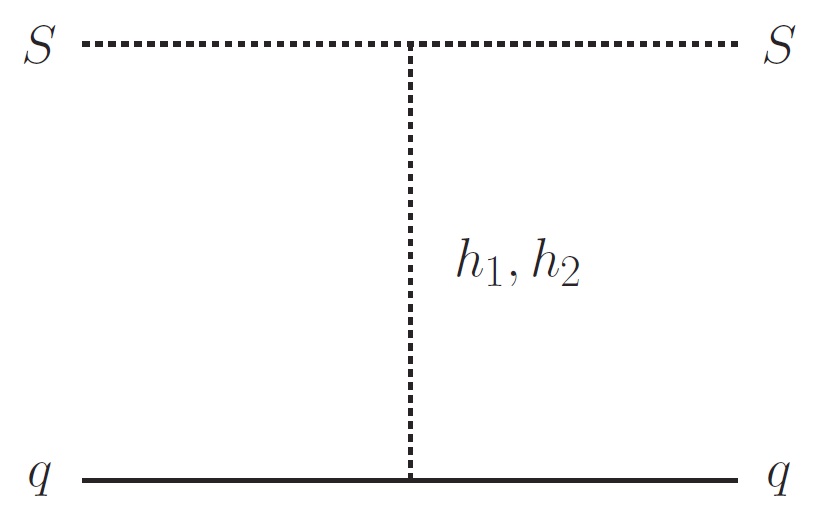}
\includegraphics[scale=1,width=7cm]{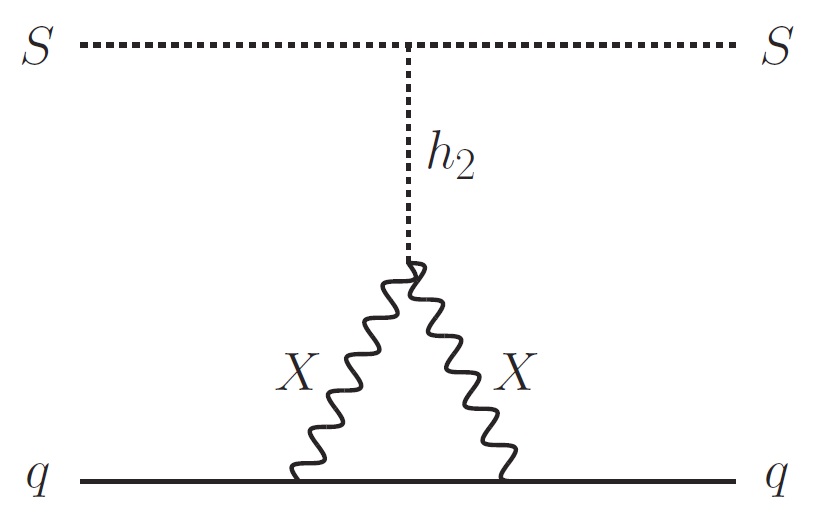}
\end{center}
\caption{Left (Right): tree (loop) diagram for DM-quark scattering.
}
\label{DD}
\end{figure}

Now we turn to the DD via DM-target nucleus scattering. The quark-level effective Lagrangian for the evaluation of the DM-nucleon spin-independent (SI) scattering cross section is given by
\begin{eqnarray}
\mathcal{L}_{eff}&=&C_q m_q SS^\ast \bar{q}q\;,
\end{eqnarray}
with
\begin{eqnarray}
C_q^{\rm tree}={s_\theta c_\theta \mu_S\over v M_2^2}-{s_\theta c_\theta \mu_S\over v M_1^2} \;, \quad
C_q^{\rm loop}\simeq {c^2_\theta \mu_S \mu_X g_q^2\over 8\pi^2 M_2^2 m_X^2}\;.
\end{eqnarray}
Besides contributions from $\phi$-Higgs couplings, here the contribution from $X$-quark coupling is also considered. The diagrams for DM-quark scattering are given in Fig.~\ref{DD}. The DM-nucleon SI scattering cross section is given by
\begin{eqnarray}
\sigma_{\rm SI}={1\over 4\pi} \Big({m_N\over m_S+m_N}\Big)^2 |C_N|^2 \;,
\end{eqnarray}
with the nucleon level coefficients being
\begin{eqnarray}
C_N&=&m_N \sum_{q=u,d,s,c,b,t}C_q f_{q}^N \;.
\end{eqnarray}
Note that the heavy quark contribution for the loop diagram is through the two-loop diagram connecting to the scalar-type DM-gluon operator~\cite{Abe:2018emu,Li:2019fnn,Ertas:2019dew} and thus can be neglected.

\begin{figure}[htbp!]
\includegraphics[width=0.45\textwidth]{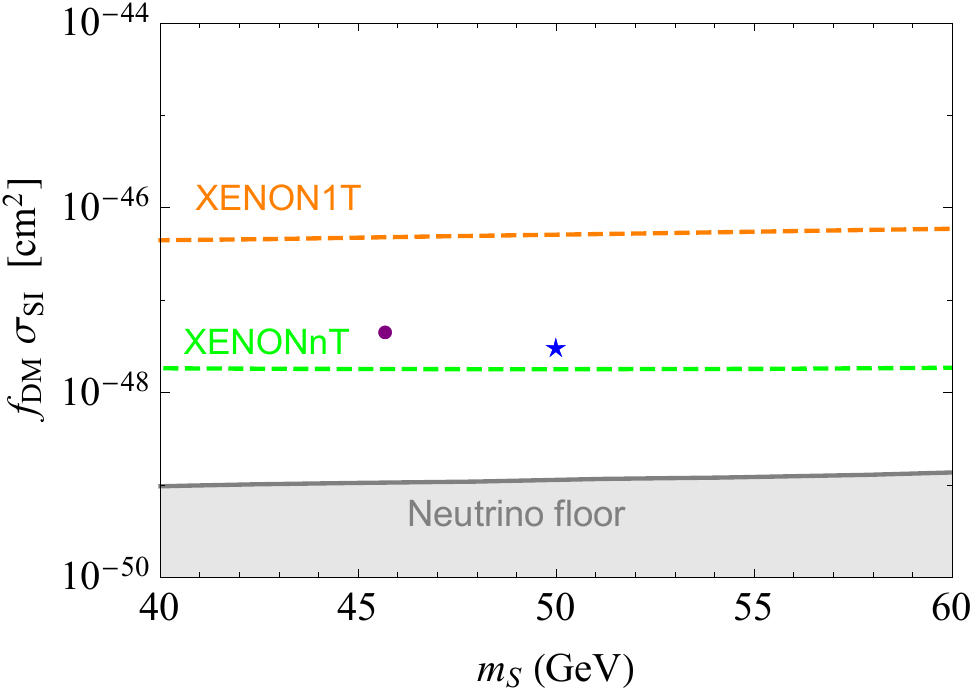} \vspace*{-1ex}
\caption{The rescaled DM-nucleon scattering cross section in the case of the tree-level contribution being dominant. Here we adopt $M_2 =$ 200 GeV and $s_\theta c_\theta \mu_S =1$ GeV. The dot (star) marker is for the WIMP (SIDM) dominant benchmark point. The upper and lower dashed curves correspond to the upper limits set by XENON1T~\cite{Aprile:2018dbl} and the projection by XENONnT (20 t$\cdot$y) \cite{Aprile:2015uzo}, respectively. The solid curve is the neutrino floor~\cite{Billard:2013qya}. }
\label{dm-dd}
\end{figure}

If the tree-level contribution is dominant in DM-nucleon scattering, the following parameter values are adopted,
\begin{eqnarray}
\sum_{q=u,d,s,c,b,t}C_q f_{q}^N = C_q^{\rm tree} f_N \;, \quad f_N = \sum_{q=u,d,s}f_{T_q}^N + {2\over 9}f_{TG}^N  \;,
\end{eqnarray}
where $f_{T_q}^N$ and $f_{TG}^N$ are light quark/gluon-nucleon form
factors, with $f_N \approx 0.308$ \cite{Hoferichter:2017olk,Alarcon:2011zs,Alarcon:2012nr}. In the case of the loop contribution from $X$-quark couplings being dominant in DM-nucleon scattering, the SI cross section induced by the loop level is
\begin{eqnarray}
\sigma_{\rm SI}&\simeq&4.3\times 10^{-64}~{\rm cm}^2 \Big({g_s\over 10^{-3}}\Big)^4 \Big({c^4_\theta \mu_S^2 \mu_X^2\over 10^{-12}~{\rm GeV}^4}\Big) \nonumber \\
&\times&\Big({100~{\rm GeV}\over M_2}\Big)^4 \Big({17~{\rm MeV}\over m_X}\Big)^4 \Big({45.7~{\rm GeV}\over m_S}\Big)^2 \; .
\end{eqnarray}
For the cross section as large as possible, here the $s$ quark's coupling parameter $g_s$ is taken to be $\sim 10^{-3}$. The typical value of this loop contribution is far below the neutrino floor. For the case that tree-level contribution dominates over the loop-level contribution in DM-nucleon scattering, the scattering cross section is shown in Fig.~\ref{dm-dd}.

The possible contribution from kinetic mixing between $X$ and $Z$ or $X$ and $\gamma$ is generated at one-loop level by virtual SM fermions is minor compared with $X$'s couplings to light quarks and electron. Thus, the possible minor contributions from kinetic mixing are neglected for simplicity.

\section{Conclusion}
\label{sec:Con}

We have investigated a small fraction of DM which mainly annihilates into a pair of on-shell vector boson $X$ followed by $X \to e^+ e^-$. The $X$ boson is indicated by $^8$Be and $^4$He anomalous transitions. The long-standing Galactic center gamma-ray excess can be explained by a small fraction of scalar DM annihilation in the cascade process of $S S^\ast \to h_1, h_2 \to X X$ followed by $X \to e^+ e^-$, with contributions from ICS emission being dominant. This DM cascade annihilation could be compatible with joint astrophysical limits from different scale astronomical observations, such as dwarf spheroidal galaxies, the Milky Way halo and galaxy groups/clusters, and meanwhile be allowed by the positron fraction observed by AMS-02. Hence, the scenario of DM annihilation is still be available to explain the Galactic center gamma-ray excess after considering these joint astrophysical limits. The loop effect from $X$-quark couplings in DM-nucleon scattering is too small to be detectable in direct detection, and the tree-level contribution is analyzed. Moreover, we suppose the couplings between $S S^\ast$ and $X$ is negligible. If there is a tiny coupling between $S S^\ast$ and $X$, the picture of DM-nucleus scattering with contact interactions fails in direct detection. A method to deal with this case was discussed in Ref.~\cite{Jia:2019qcs}. We look forward to the further investigation of DM at the future joint detections.

\acknowledgments

T.~Li would like to thank Jia Liu for very useful discussion and communication. The work of T.~Li was supported by the the National Natural Science Foundation of China (Grant No. 11975129, 12035008) and ``the Fundamental Research Funds for the Central Universities'', Nankai University (Grant No. 63196013). L.-B. Jia was supported by the Longshan academic talent research supporting program of SWUST under Contract No. 18LZX415.



\begin{thebibliography}{0}


\bibitem{Cui:2017nnn}
  X.~Cui {\it et al.} [PandaX-II Collaboration],
  Phys.\ Rev.\ Lett.\  {\bf 119}, no. 18, 181302 (2017)
  [arXiv:1708.06917 [astro-ph.CO]].


\bibitem{Akerib:2016vxi}
  D.~S.~Akerib {\it et al.} [LUX Collaboration],
  Phys.\ Rev.\ Lett.\  {\bf 118}, no. 2, 021303 (2017)
  [arXiv:1608.07648 [astro-ph.CO]].


\bibitem{Aprile:2018dbl}
  E.~Aprile {\it et al.} [XENON Collaboration],
  Phys.\ Rev.\ Lett.\  {\bf 121}, no. 11, 111302 (2018)
  [arXiv:1805.12562 [astro-ph.CO]].


\bibitem{Akerib:2017kat}
  D.~S.~Akerib {\it et al.} [LUX Collaboration],
  Phys.\ Rev.\ Lett.\  {\bf 118}, no. 25, 251302 (2017)
  [arXiv:1705.03380 [astro-ph.CO]].


\bibitem{Xia:2018qgs}
  J.~Xia {\it et al.} [PandaX-II Collaboration],
  Phys.\ Lett.\ B {\bf 792}, 193 (2019)
  [arXiv:1807.01936 [hep-ex]].


\bibitem{Aprile:2019dbj}
  E.~Aprile {\it et al.} [XENON Collaboration],
  Phys.\ Rev.\ Lett.\  {\bf 122}, no. 14, 141301 (2019)
  [arXiv:1902.03234 [astro-ph.CO]].


\bibitem{Amole:2019fdf}
  C.~Amole {\it et al.} [PICO Collaboration],
  Phys.\ Rev.\ D {\bf 100}, no. 2, 022001 (2019)
  [arXiv:1902.04031 [astro-ph.CO]].



\bibitem{Goodenough:2009gk}
  L.~Goodenough and D.~Hooper,
  arXiv:0910.2998 [hep-ph].


\bibitem{Hooper:2010mq}
  D.~Hooper and L.~Goodenough,
  Phys.\ Lett.\ B {\bf 697}, 412 (2011)
  [arXiv:1010.2752 [hep-ph]].


\bibitem{Hooper:2011ti}
  D.~Hooper and T.~Linden,
  Phys.\ Rev.\ D {\bf 84}, 123005 (2011)
  [arXiv:1110.0006 [astro-ph.HE]].


\bibitem{Abazajian:2012pn}
  K.~N.~Abazajian and M.~Kaplinghat,
  Phys.\ Rev.\ D {\bf 86}, 083511 (2012)
  Erratum: [Phys.\ Rev.\ D {\bf 87}, 129902 (2013)]
  [arXiv:1207.6047 [astro-ph.HE]].


\bibitem{Gordon:2013vta}
  C.~Gordon and O.~Macias,
  Phys.\ Rev.\ D {\bf 88}, no. 8, 083521 (2013)
  Erratum: [Phys.\ Rev.\ D {\bf 89}, no. 4, 049901 (2014)]
  [arXiv:1306.5725 [astro-ph.HE]].


\bibitem{Daylan:2014rsa}
  T.~Daylan, D.~P.~Finkbeiner, D.~Hooper, T.~Linden, S.~K.~N.~Portillo, N.~L.~Rodd and T.~R.~Slatyer,
  Phys.\ Dark Univ.\  {\bf 12}, 1 (2016)
  [arXiv:1402.6703 [astro-ph.HE]].


\bibitem{Calore:2014nla}
  F.~Calore, I.~Cholis, C.~McCabe and C.~Weniger,
  Phys.\ Rev.\ D {\bf 91}, no. 6, 063003 (2015)
  [arXiv:1411.4647 [hep-ph]].


\bibitem{Hooper:2013nhl}
  D.~Hooper, I.~Cholis, T.~Linden, J.~Siegal-Gaskins and T.~Slatyer,
  Phys.\ Rev.\ D {\bf 88}, 083009 (2013)
  [arXiv:1305.0830 [astro-ph.HE]].


\bibitem{Yuan:2014rca}
  Q.~Yuan and B.~Zhang,
  JHEAp {\bf 3-4}, 1 (2014)
  [arXiv:1404.2318 [astro-ph.HE]].


\bibitem{Cholis:2014lta}
  I.~Cholis, D.~Hooper and T.~Linden,
  JCAP {\bf 1506}, 043 (2015)
  [arXiv:1407.5625 [astro-ph.HE]].


\bibitem{Petrovic:2014xra}
  J.~Petrovi$\acute{c}$, P.~D.~Serpico and G.~Zaharijas,
  JCAP {\bf 1502}, 023 (2015)
  [arXiv:1411.2980 [astro-ph.HE]].


\bibitem{Bartels:2015aea}
  R.~Bartels, S.~Krishnamurthy and C.~Weniger,
  Phys.\ Rev.\ Lett.\  {\bf 116}, no. 5, 051102 (2016)
  [arXiv:1506.05104 [astro-ph.HE]].


\bibitem{Lee:2015fea}
  S.~K.~Lee, M.~Lisanti, B.~R.~Safdi, T.~R.~Slatyer and W.~Xue,
  Phys.\ Rev.\ Lett.\  {\bf 116}, no. 5, 051103 (2016)
  [arXiv:1506.05124 [astro-ph.HE]].


\bibitem{Buschmann:2020adf}
 M.~Buschmann, N.~L.~Rodd, B.~R.~Safdi, L.~J.~Chang, S.~Mishra-Sharma, M.~Lisanti and O.~Macias,
Phys. Rev. D \textbf{102}, no.2, 023023 (2020)
[arXiv:2002.12373 [astro-ph.HE]].


\bibitem{Leane:2019xiy}
  R.~K.~Leane and T.~R.~Slatyer,
  Phys.\ Rev.\ Lett.\  {\bf 123}, no. 24, 241101 (2019)
  [arXiv:1904.08430 [astro-ph.HE]].


\bibitem{Zhong:2019ycb}
Y.~M.~Zhong, S.~D.~McDermott, I.~Cholis and P.~J.~Fox,
Phys. Rev. Lett. \textbf{124}, no.23, 231103 (2020)
[arXiv:1911.12369 [astro-ph.HE]].


\bibitem{Leane:2020nmi}
R.~K.~Leane and T.~R.~Slatyer,
Phys. Rev. Lett. \textbf{125}, no.12, 121105 (2020)
[arXiv:2002.12370 [astro-ph.HE]].


\bibitem{Leane:2020pfc}
R.~K.~Leane and T.~R.~Slatyer,
Phys. Rev. D \textbf{102}, no.6, 063019 (2020)
[arXiv:2002.12371 [astro-ph.HE]].


\bibitem{Hoof:2018hyn}
  S.~Hoof, A.~Geringer-Sameth and R.~Trotta,
  JCAP {\bf 2002}, 012 (2020)
  [arXiv:1812.06986 [astro-ph.CO]].


\bibitem{Chang:2018bpt}
  L.~J.~Chang, M.~Lisanti and S.~Mishra-Sharma,
  Phys.\ Rev.\ D {\bf 98}, no. 12, 123004 (2018)
  [arXiv:1804.04132 [astro-ph.CO]].


\bibitem{Abazajian:2020tww}
K.~N.~Abazajian, S.~Horiuchi, M.~Kaplinghat, R.~E.~Keeley and O.~Macias,
Phys. Rev. D \textbf{102}, no.4, 043012 (2020)
[arXiv:2003.10416 [hep-ph]].


\bibitem{Lisanti:2017qlb}
  M.~Lisanti, S.~Mishra-Sharma, N.~L.~Rodd and B.~R.~Safdi,
  Phys.\ Rev.\ Lett.\  {\bf 120}, no. 10, 101101 (2018)
  [arXiv:1708.09385 [astro-ph.CO]].


\bibitem{Lisanti:2017qoz}
  M.~Lisanti, S.~Mishra-Sharma, N.~L.~Rodd, B.~R.~Safdi and R.~H.~Wechsler,
  Phys.\ Rev.\ D {\bf 97}, no. 6, 063005 (2018)
  [arXiv:1709.00416 [astro-ph.CO]].


\bibitem{Tan:2019gmb}
X.~Tan, M.~Colavincenzo and S.~Ammazzalorso,
Mon. Not. Roy. Astron. Soc. \textbf{495}, no.1, 114-122 (2020)
[arXiv:1907.06905 [astro-ph.CO]].


\bibitem{Liu:2014cma}
J.~Liu, N.~Weiner and W.~Xue,
JHEP \textbf{08}, 050 (2015)
[arXiv:1412.1485 [hep-ph]].


\bibitem{Kaplinghat:2015gha}
M.~Kaplinghat, T.~Linden and H.~Yu,
Phys.\ Rev.\ Lett.\  \textbf{114}, no.21, 211303 (2015)
[arXiv:1501.03507 [hep-ph]].


\bibitem{Krasznahorkay:2015iga}
  A.~J.~Krasznahorkay {\it et al.},
  Phys.\ Rev.\ Lett.\  {\bf 116}, no. 4, 042501 (2016)
  [arXiv:1504.01527 [nucl-ex]].


\bibitem{Krasznahorkay:2019lyl}
  A.~J.~Krasznahorkay {\it et al.},
  arXiv:1910.10459 [nucl-ex].


\bibitem{Kozaczuk:2016nma}
  J.~Kozaczuk, D.~E.~Morrissey and S.~R.~Stroberg,
  Phys.\ Rev.\ D {\bf 95}, no. 11, 115024 (2017)
  [arXiv:1612.01525 [hep-ph]].


\bibitem{Sommerfeld:1931}
  A.~Sommerfeld,
  Annalen der Physik {\bf 403}, 257 (1931).


\bibitem{Zurek:2008qg}
  K.~M.~Zurek,
  Phys.\ Rev.\ D {\bf 79}, 115002 (2009)
  [arXiv:0811.4429 [hep-ph]].


\bibitem{Profumo:2009tb}
  S.~Profumo, K.~Sigurdson and L.~Ubaldi,
  JCAP {\bf 0912}, 016 (2009)
  [arXiv:0907.4374 [hep-ph]].


\bibitem{Zhang:2017zap}
  X.~Zhang and G.~A.~Miller,
  Phys.\ Lett.\ B {\bf 773}, 159 (2017)
  [arXiv:1703.04588 [nucl-th]].


\bibitem{Feng:2016jff}
  J.~L.~Feng, B.~Fornal, I.~Galon, S.~Gardner, J.~Smolinsky, T.~M.~P.~Tait and P.~Tanedo,
  Phys.\ Rev.\ Lett.\  {\bf 117}, no. 7, 071803 (2016)
  [arXiv:1604.07411 [hep-ph]].


\bibitem{Feng:2016ysn}
  J.~L.~Feng, B.~Fornal, I.~Galon, S.~Gardner, J.~Smolinsky, T.~M.~P.~Tait and P.~Tanedo,
  Phys.\ Rev.\ D {\bf 95}, no. 3, 035017 (2017)
  [arXiv:1608.03591 [hep-ph]].


\bibitem{Gu:2016ege}
  P.~H.~Gu and X.~G.~He,
  Nucl.\ Phys.\ B {\bf 919}, 209 (2017)
  [arXiv:1606.05171 [hep-ph]].


\bibitem{Feng:2020mbt}
J.~L.~Feng, T.~M.~P.~Tait and C.~B.~Verhaaren,
Phys. Rev. D \textbf{102}, no.3, 036016 (2020)
[arXiv:2006.01151 [hep-ph]].





\bibitem{Seto:2020jal}
  O.~Seto and T.~Shimomura,
  arXiv:2006.05497 [hep-ph].


\bibitem{Ellwanger:2016wfe}
  U.~Ellwanger and S.~Moretti,
  JHEP {\bf 1611}, 039 (2016)
  [arXiv:1609.01669 [hep-ph]].





\bibitem{Liang:2016ffe}
  Y.~Liang, L.~B.~Chen and C.~F.~Qiao,
  Chin.\ Phys.\ C {\bf 41}, no. 6, 063105 (2017)
  [arXiv:1607.08309 [hep-ph]].


\bibitem{Fayet:2016nyc}
  P.~Fayet,
  Eur.\ Phys.\ J.\ C {\bf 77}, no. 1, 53 (2017)
  [arXiv:1611.05357 [hep-ph]].


\bibitem{DelleRose:2017xil}
  L.~Delle Rose, S.~Khalil and S.~Moretti,
  Phys.\ Rev.\ D {\bf 96}, no. 11, 115024 (2017)
  [arXiv:1704.03436 [hep-ph]].


\bibitem{Fornal:2017msy}
  B.~Fornal,
  Int.\ J.\ Mod.\ Phys.\ A {\bf 32}, 1730020 (2017)
  [arXiv:1707.09749 [hep-ph]].


\bibitem{Jiang:2018uhs}
  J.~Jiang, L.~B.~Chen, Y.~Liang and C.~F.~Qiao,
  Eur.\ Phys.\ J.\ C {\bf 78}, no. 6, 456 (2018).


\bibitem{Nam:2019osu}
C.~H.~Nam,
Eur. Phys. J. C \textbf{80}, no.3, 231 (2020)
[arXiv:1907.09819 [hep-ph]].



\bibitem{Riordan:1987aw}
E.~M.~Riordan, M.~W.~Krasny, K.~Lang, P.~De Barbaro, A.~Bodek, S.~Dasu, N.~Varelas, X.~Wang, R.~G.~Arnold and D.~Benton, \textit{et al.}
Phys. Rev. Lett. \textbf{59}, 755 (1987)


\bibitem{Alexander:2016aln}
J.~Alexander, M.~Battaglieri, B.~Echenard, R.~Essig, M.~Graham, E.~Izaguirre, J.~Jaros, G.~Krnjaic, J.~Mardon and D.~Morrissey, \textit{et al.}
[arXiv:1608.08632 [hep-ph]].


\bibitem{Banerjee:2019hmi}
D.~Banerjee \textit{et al.} [NA64],
Phys. Rev. D \textbf{101}, no.7, 071101 (2020)
[arXiv:1912.11389 [hep-ex]].


\bibitem{Arvieux:1967tqs}
J.~Arvieux, P.~Darriulat, D.~Garreta, A.~Papineau, A.~Tarrats and J.~Testoni,
Nucl. Phys. A \textbf{94}, 663-672 (1967)
doi:10.1016/0375-9474(67)90439-3

\bibitem{Subotic:1978mab}
K.~M.~Suboti\'c, B.~Lalovi\'c and B.~Z.~Stepan\v{c}i\'c,
Nucl. Phys. A \textbf{296}, 141-150 (1978)
doi:10.1016/0375-9474(78)90417-7

\bibitem{Balewski:2014pxa}
J.~Balewski, J.~Bernauer, J.~Bessuille, R.~Corliss, R.~Cowan, C.~Epstein, P.~Fisher, D.~Hasell, E.~Ihloff and Y.~Kahn, \textit{et al.}
[arXiv:1412.4717 [physics.ins-det]].


\bibitem{Moreno:2013mja}
O.~Moreno,
[arXiv:1310.2060 [physics.ins-det]].















\bibitem{Jia:2016uxs}
  L.~B.~Jia and X.~Q.~Li,
  Eur.\ Phys.\ J.\ C {\bf 76}, no. 12, 706 (2016)
  [arXiv:1608.05443 [hep-ph]].


\bibitem{Kitahara:2016zyb}
  T.~Kitahara and Y.~Yamamoto,
  Phys.\ Rev.\ D {\bf 95}, no. 1, 015008 (2017)
  [arXiv:1609.01605 [hep-ph]].


\bibitem{Chen:2016tdz}
  C.~S.~Chen, G.~L.~Lin, Y.~H.~Lin and F.~Xu,
  Int.\ J.\ Mod.\ Phys.\ A {\bf 32}, no. 31, 1750178 (2017)
  [arXiv:1609.07198 [hep-ph]].


\bibitem{Seto:2016pks}
  O.~Seto and T.~Shimomura,
  Phys.\ Rev.\ D {\bf 95}, no. 9, 095032 (2017)
  [arXiv:1610.08112 [hep-ph]].


\bibitem{Jia:2017iyc}
  L.~B.~Jia,
  Eur.\ Phys.\ J.\ C {\bf 78}, no. 2, 112 (2018)
  [arXiv:1710.03906 [hep-ph]].


\bibitem{Jia:2018mkc}
  L.~B.~Jia, X.~J.~Deng and C.~F.~Liu,
  Eur.\ Phys.\ J.\ C {\bf 78}, no. 11, 956 (2018)
  [arXiv:1809.00177 [hep-ph]].


\bibitem{Pospelov:2007mp}
M.~Pospelov, A.~Ritz and M.~B.~Voloshin,
Phys. Lett. B \textbf{662}, 53-61 (2008)
[arXiv:0711.4866 [hep-ph]].


\bibitem{OConnell:2006rsp}
D.~O'Connell, M.~J.~Ramsey-Musolf and M.~B.~Wise,
Phys. Rev. D \textbf{75}, 037701 (2007)
[arXiv:hep-ph/0611014 [hep-ph]].


\bibitem{Steigman:2012nb}
  G.~Steigman, B.~Dasgupta and J.~F.~Beacom,
  Phys.\ Rev.\ D {\bf 86}, 023506 (2012)
  [arXiv:1204.3622 [hep-ph]].


\bibitem{Kaplinghat:2013xca}
  M.~Kaplinghat, R.~E.~Keeley, T.~Linden and H.~B.~Yu,
  Phys.\ Rev.\ Lett.\  {\bf 113}, 021302 (2014)
  [arXiv:1311.6524 [astro-ph.CO]].


\bibitem{Jia:2019yhr}
  L.~B.~Jia and X.~Liao,
  Phys.\ Rev.\ D {\bf 100}, no. 3, 035012 (2019)
  [arXiv:1906.00559 [hep-ph]].


\bibitem{Elor:2015bho}
  G.~Elor, N.~L.~Rodd, T.~R.~Slatyer and W.~Xue,
  JCAP {\bf 1606}, 024 (2016)
  [arXiv:1511.08787 [hep-ph]].


\bibitem{Abe:2018emu}
T.~Abe, M.~Fujiwara and J.~Hisano,
JHEP \textbf{02}, 028 (2019)
[arXiv:1810.01039 [hep-ph]].


\bibitem{Li:2019fnn}
T.~Li and P.~Wu,
Chin. Phys. C \textbf{43}, no.11, 113102 (2019)
[arXiv:1904.03407 [hep-ph]].


\bibitem{Ertas:2019dew}
F.~Ertas and F.~Kahlhoefer,
JHEP \textbf{06}, 052 (2019)
[arXiv:1902.11070 [hep-ph]].


\bibitem{Hoferichter:2017olk}
  M.~Hoferichter, P.~Klos, J.~Men$\acute{e}$ndez and A.~Schwenk,
  Phys.\ Rev.\ Lett.\  {\bf 119}, no. 18, 181803 (2017)
  [arXiv:1708.02245 [hep-ph]].


\bibitem{Alarcon:2011zs}
J.~M.~Alarcon, J.~Martin Camalich and J.~A.~Oller,
Phys. Rev. D \textbf{85}, 051503 (2012)
[arXiv:1110.3797 [hep-ph]].


\bibitem{Alarcon:2012nr}
J.~M.~Alarcon, L.~S.~Geng, J.~Martin Camalich and J.~A.~Oller,
Phys. Lett. B \textbf{730}, 342-346 (2014)
[arXiv:1209.2870 [hep-ph]].


\bibitem{Aprile:2015uzo}
  E.~Aprile {\it et al.} [XENON Collaboration],
  JCAP {\bf 1604}, 027 (2016)
  [arXiv:1512.07501 [physics.ins-det]].


\bibitem{Billard:2013qya}
  J.~Billard, L.~Strigari and E.~Figueroa-Feliciano,
  Phys.\ Rev.\ D {\bf 89}, no. 2, 023524 (2014)
  [arXiv:1307.5458 [hep-ph]].


\bibitem{Jia:2019qcs}
  L.~B.~Jia,
  Eur.\ Phys.\ J.\ C {\bf 80}, no. 2, 143 (2020)
  [arXiv:1910.05633 [hep-ph]].






\end{thebibliography}
\end{document}